\documentclass[prb,twocolumn,showpacs,preprintnumbers,amsmath,amssymb]{revtex4-1}

\usepackage{graphicx,rotating}
\usepackage{dcolumn}
\usepackage{bm}
\usepackage{epsfig}
\usepackage{longtable}
\usepackage{color}

\begin{document}

\title{Contacts for organic switches with carbon-nanotube leads.}

\author{Ma\l{}gorzata~Wierzbowska\email{wierzbowska@ifpan.edu.pl},
Micha\l{} F. Rode, Miko\l{}aj Sadek and Andrzej~L.~Sobolewski}

\affiliation{Institut of Physics, Polish Academy of Science (PAS),
Al. Lotnik\'ow 32/46, 02-668 Warszawa, Poland}

\date{\today}

\begin{abstract}
We focus on two classes of organic switches operating due to photo- or field-induced 
proton transfer (PT) process. By means of the first-principles simulations,
we search for the atomic contacts which strengthen diversity of the two swapped I-V
characteristics between two tautomers. We emphasize, that the low-resistive contacts 
not necessarily possess good switching properties. Very often, the higher-current flow
makes it more difficult to distinguish between the logic states. Instead,
the more resistive contacts multiply a current gear to a larger extent.
The low- and high-bias work regimes set additional conditions, 
which are fulfilled by different
contacts: (i) in the very low-voltage regime, the direct connections to the nanotubes 
perform better than the popular sulfur contacts,
(ii) in the higher-voltage regime, the best are the peroxide (-O-O-) contacts.
Additionally, we find that the switching-bias value is not an inherent property of 
the conducting molecule, but it strongly depends on the chosen contacts.
\end{abstract}

\pacs {73.23.Ad, 73.63.Rt, 73.63.Fg, 71.15.Mb}

\maketitle

\section{Introduction} 

Miniaturization caused mainly by the energy savings, 
speed and multi-functionality reasons is a strong driving force for the molecular
basic research. The diodes, transistors, switches, memory cells, and sensors
are currently designed from the molecular systems connected to
the metallic or carbon leads.\cite{b1} 
Fabrication of organic devices is technologically simpler and its
costs are lower. Thus, the search for organic substitutes of inorganic 
systems is a vigorous branch.\cite{b2}
It is also dictated by the ecologic requirements since in the organic systems, 
the light atoms replace heavy - and toxic for the environment - elements.
Moreover, the relatively new field of molecular electronics
intensively stimulates the development of methods.\cite{b3}

The metallic leads are widely used for nano-device electronics,\cite{b4} 
however recently, the carbon systems become more popular.\cite{b5} This is mainly due
to the better thermoelectric efficiency - which is a separate subject of
investigations including the vibronic effects.\cite{b6,b7,b8,b9} 
Therefore, we decided to use the carbon-nanotube (CNT) leads.
To date, the carbon-system contacts to the metal leads were explored for C60,\cite{b10,b11} 
CNT\cite{b12,b13} and graphene.\cite{b13,b14}

Various types of electrodes and nano-devices, as well as their functionalities, 
set requirements for the electronic contacts.\cite{b15,b16}
Their role has been addressed a decade ago,\cite{b17,b18} and
it is still of major importance.\cite{b19,b3}
All types of devices seek the low-resistivity contacts
in order to avoid large heat dissipation.\cite{b20,b21,b22} Additionally, some
contacts are designed to be spin-selective.\cite{b23} Switches, however, 
need contacts which are low-resistive, and additionally they amplify 
the current gear obtained 
at the transition between two, or more, molecular forms; so
an effect of the switched logic state is easily measured in the circuit. 
Finding such contacts is the main goal of
our study, since – according to our knowledge – it was not addressed up to date.

Molecular switches work under various external triggering stimuli: 
electric or magnetic field, light, temperature, chemical reaction, mechanical stress,
radiation, etc.\cite{b24} 
Here, we focus on two classes of the molecular switches, which operate under the
photo- or field-stimuli. The first group can operate at very low voltage, 
since the tautomeric transition is triggered by light, not by bias.
The second group works in the wide range of voltages,
set by the values of swapping biases – which are different for the state-on and the state-off. 
These voltages typically are higher than the low-bias regime in which 
the photo-switches operate.
The higher-bias regime is more demanding for the electronic contacts.
Hence, it is much more difficult to obtain a high current gear
for the field-switches than for the photo-switches.

In this work, we used the first-principles methods for the electronic-structure and 
ballistic-conductance calculations to obtain the current-voltage (I-V) curves
for the tautomeric switches connected to the CNT leads.
The studied photo- and field-switches base on the proton transfer between the oxygen 
and the nitrogen centers.\cite{b25,b26,b27,b28,b29,b30}
The contacts are chosen among the light atoms or chemical groups: 
-C-, -C$\equiv$C-, -O-, -O-O-, -S-, -CS$_2$-, -CH$_{2}$-, -NH-,
as well as the direct connection to the CNT.
Some choices might appear fancy at the first sight,
however the results on a nanoscale are often surprising.\cite{b31}

Interestingly, the switching voltages are not 
the properties of the conducting molecules only, but they depend strongly
on the selected contacts.
Among the investigated in this work structures, the cases with multiple negative differential 
resistance (NDR) show up, and they were described in our previous work as candidates
for the high-frequency devices.\cite{b32}

\section{Results and discussion}

It is a common practice that the sulfur contacts are used to connect a molecule to the 
metallic or carbon-based leads.
Some authors connect the molecule directly to the graphene surface or the carbon
nanotube.\cite{b33} Here, we check some other possibilities.  
Among others, we explore the properties of acetylene (-C$\equiv$C-), oxygen (-O-) and
peroxide (-O-O-) contacts. Generally, the oxygen element is not popular 
as a contact. But when oxygen is added to some conductive structures, 
it can bring along a new physics.\cite{b34,b35,b36}
Below, we report the results for the two types of molecular switches: 
triggered either by light or by voltage; both are based on the proton transfer (PT)
reaction along the intramolecular hydrogen-bond.

\subsection{Photo-switches: low-voltage work regime}

\begin{figure}[ht]
\includegraphics[scale=0.25,angle=0.0]{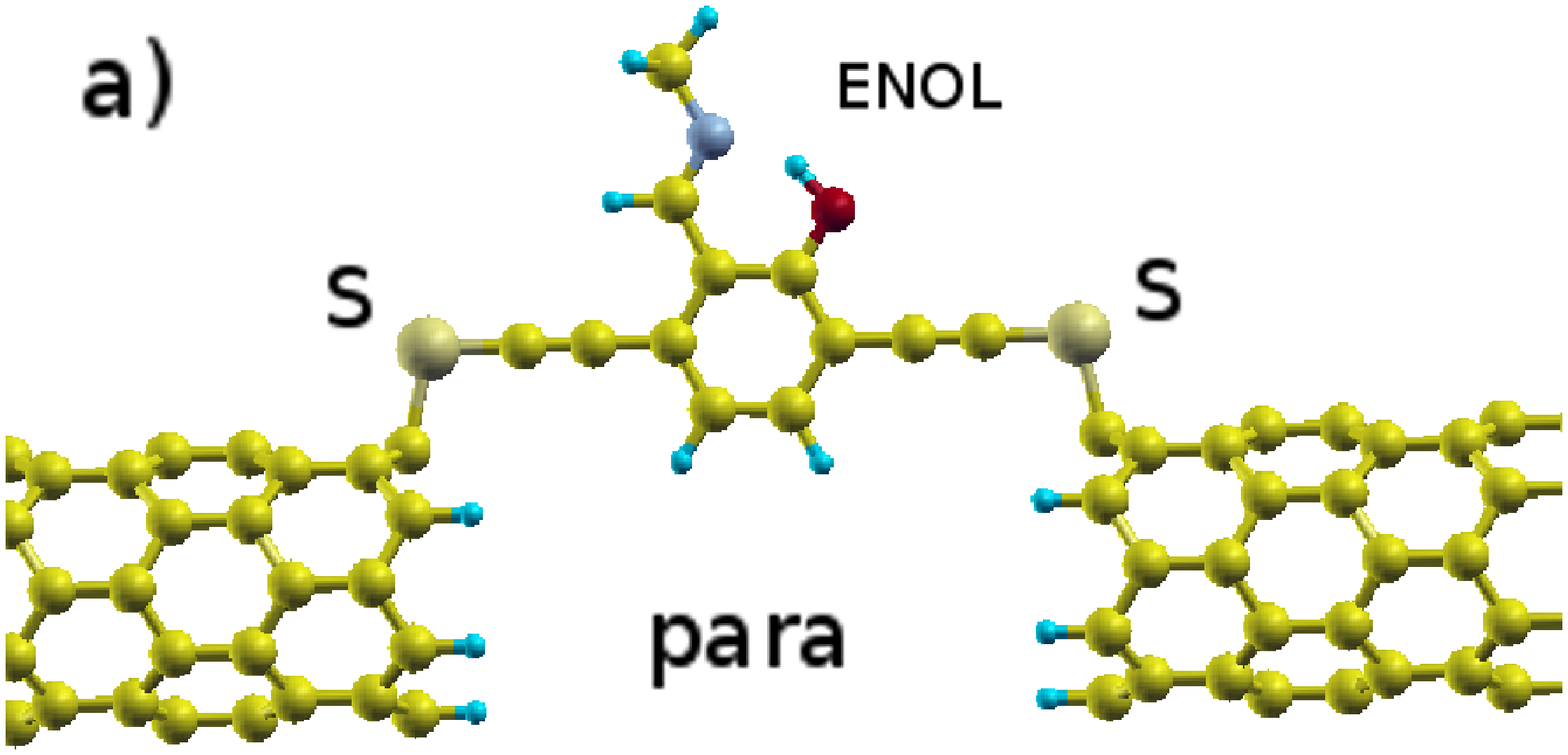}
\includegraphics[scale=0.25,angle=0.0]{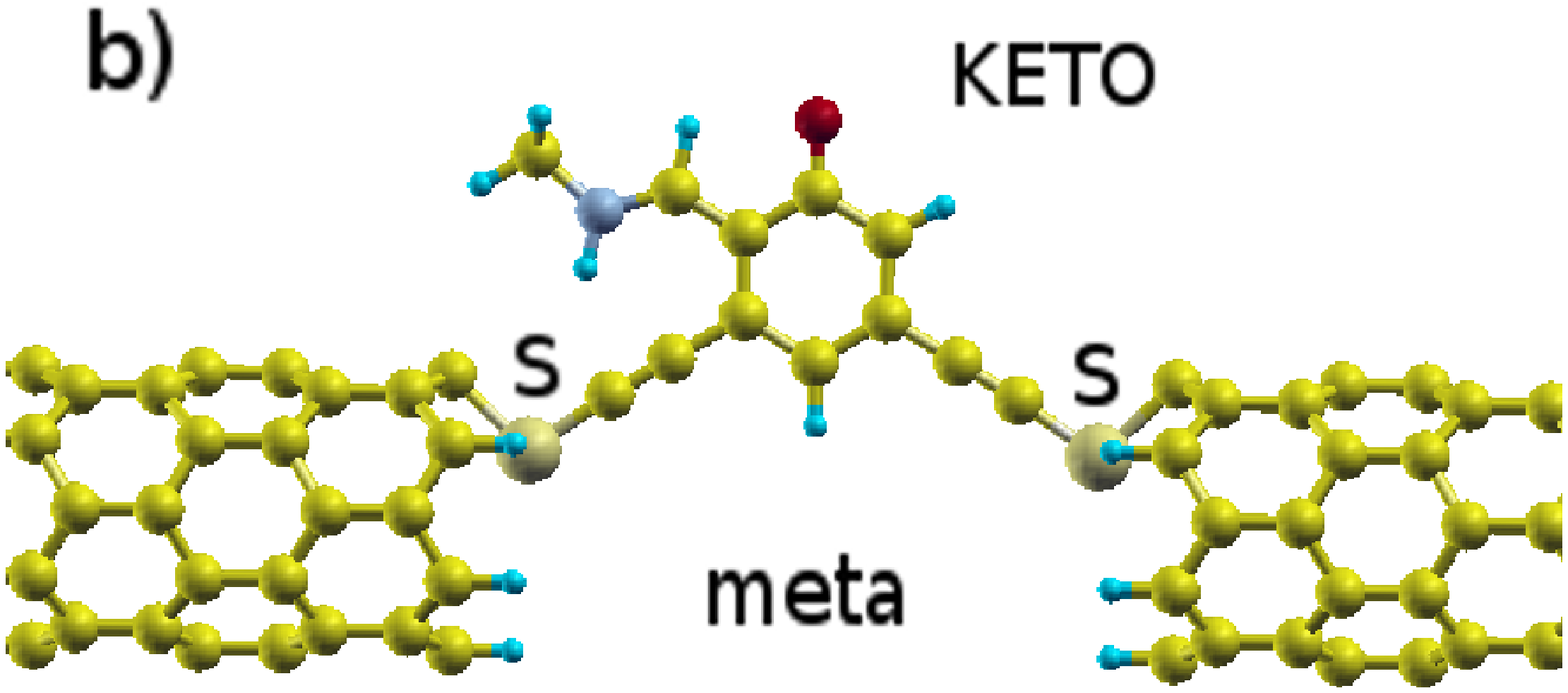}
\includegraphics[scale=0.21,angle=0.0]{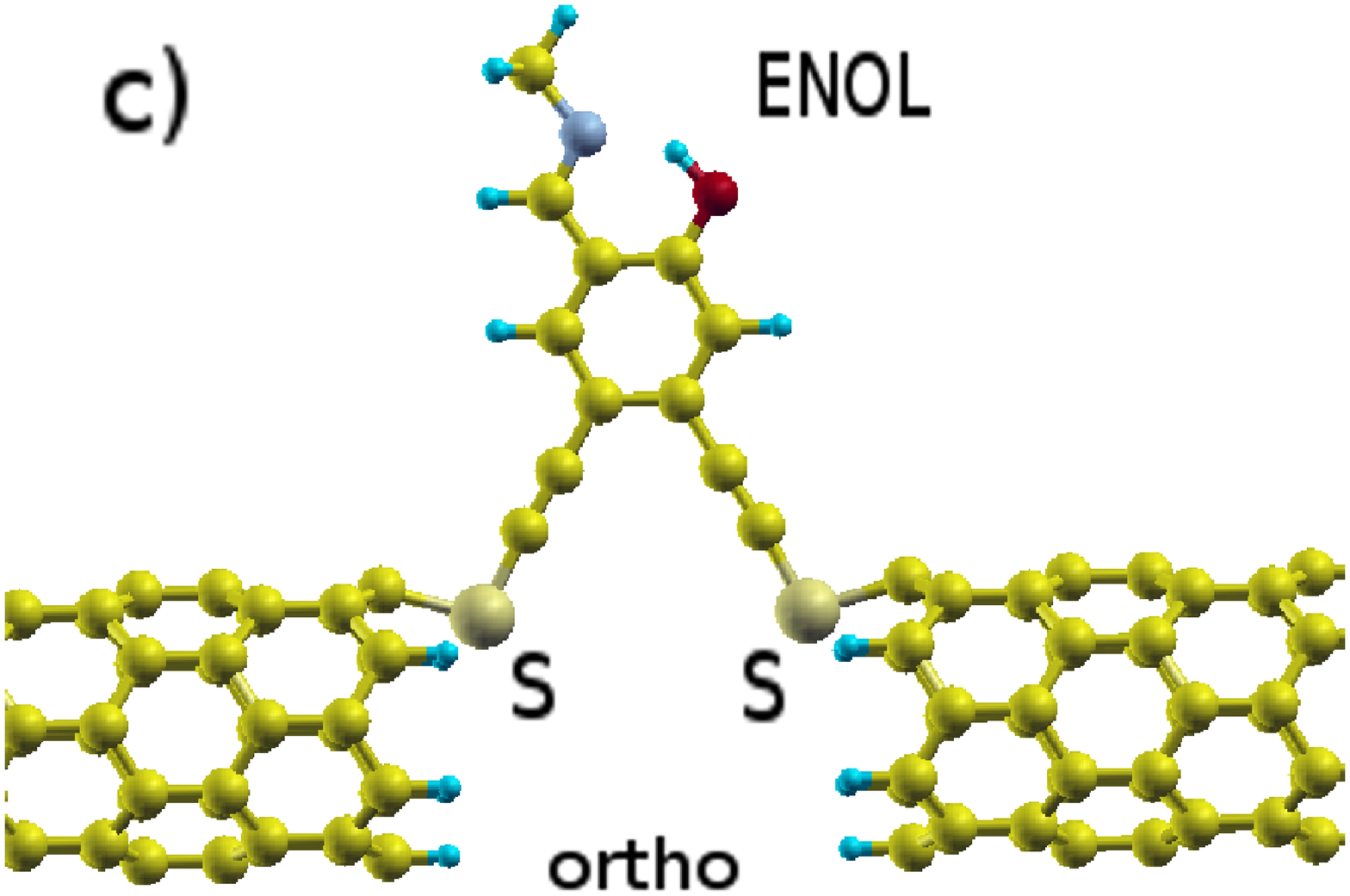}
\caption{Tautomeric forms and geometries of the photo-switches
in the (a) {\it -para}, (b) {\it -meta} and (c) {\it -ortho} configuration.}
\end{figure}

The photo-switches discussed in this subsection, differ from the discussed hereafter 
field-switches by the fact, that their two photo-tautomeric structures form 
due to reversible photoinduced intramolecular proton-transfer process  
- while the field-switching induces this reaction in the ground electronic state.
Photo-switching of molecular conductance is studied for the two stable isomers 
of salicylidene methylamine (SMA).\cite{b27,b28}
The photo-induced PT reaction occurs between the oxygen atom of 
the hydroxyl group and the
nitrogen atom of the amino group of SMA.\cite{b30} 
Both respective photo-tautomeric forms - called the enol and keto -  are
presented in Fig.~1. Additionally, these forms differ 
from each other by the conformation of the methylamine tail
with respect to the salicylidene core.\cite{b27}
Importance of the contact geometry for the transport properties              
has been shown for many systems.\cite{b10,b37,b38,b39,b40}
Moreover, the properties of the junction may also depend on
the CNT chirality, thus, all carbon electrodes considered in this work are of           
the zigzag type - the most effective form for the switches.\cite{b41}

Following the work by Staykov et al.,\cite{b30} we simulate transport in various 
structural connections to the leads:
{\it para}, {\it meta}, and not studied earlier {\it ortho} - all displayed in Fig.~1. 
But our main aim is to check the switching impact of contacts,
such as the direct connection, O and S atomic connections. The S and O connections
form the angular structures with the CNT (about 105-120 deg). 
In contrast, the direct connection is linear. 
Additionally, the electronic conjugation of the phenyl ring of SMA is
extended by acetylenic moieties attached on its both sides; so-called anchor groups.
Former studies of these molecules framed, however, between the metallic electrodes, 
have revealed that the {\it meta} connection is much more efficient for the photo-switching 
than the {\it para} one.\cite{b30}
In this work, we considered also the {\it ortho} structure.
The obtained I-V curves are presented in Fig.~2 (details of the calculations
are given in the methods section).

\begin{figure}[ht]
\includegraphics[scale=0.4,angle=0.0]{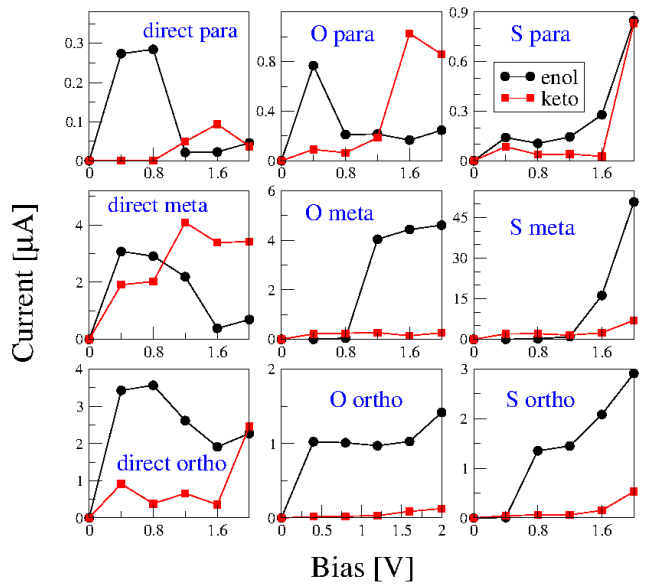}
\caption{The I-V characteristics of the photo-switches - in the enol (black) and keto (red) forms
- with various contacts: direct (molecule-CNT), or via oxygen and sulfur;
for the three geometric connections: {\it para}, {\it meta} and {\it ortho}.}
\end{figure}

Focusing on the resistivity of the contacts and the connecting structures, 
generally it is not surprising that the sulfur atom is the best conductive contact within the
range of applied bias (about 2V). 
The {\it meta} structures are much more conductive than the {\it para}, 
and also more than the {\it ortho}. With an exception that for the direct connection,
the {\it meta} and {\it ortho} geometries show similar currents for 
the enol and keto structures below 2V.
Transition voltages, in a sense of the voltage at which the
current rises from zero for very small applied biases,\cite{b42} are usually lower for 
-O- than -S-, and are the lowest for the direct connection.
This fact sets the working bias-regimes:
V$_{work}$(direct)~$<$~V$_{work}$(O)~$<$~V$_{work}$(S).

Further inspection of the results presented in Fig.~2 leads to the following conclusions: 
(i) the most interesting result with respect to the current gain between the two isomeric forms
at very low bias, up to 0.4V, is obtained for the cases: 
direct- and oxygen {\it para} and {\it ortho}, 
but it is not the case with the S contact, (ii) for the voltage
around 2V, the best performing switches are: oxygen contacts and direct-{\it meta}, 
and S-{\it meta} and {\it -ortho}. 
For some cases, the negative differential resistance is observed - especially with 
the direct contacts.
This effect, however, does not exclude the device from operation 
if we choose the proper working bias-range ($<$0.4-0.8V).

\subsection{Field-switches: higher-voltage work regime}

\begin{figure}[ht]
\includegraphics[scale=0.19,angle=0.0]{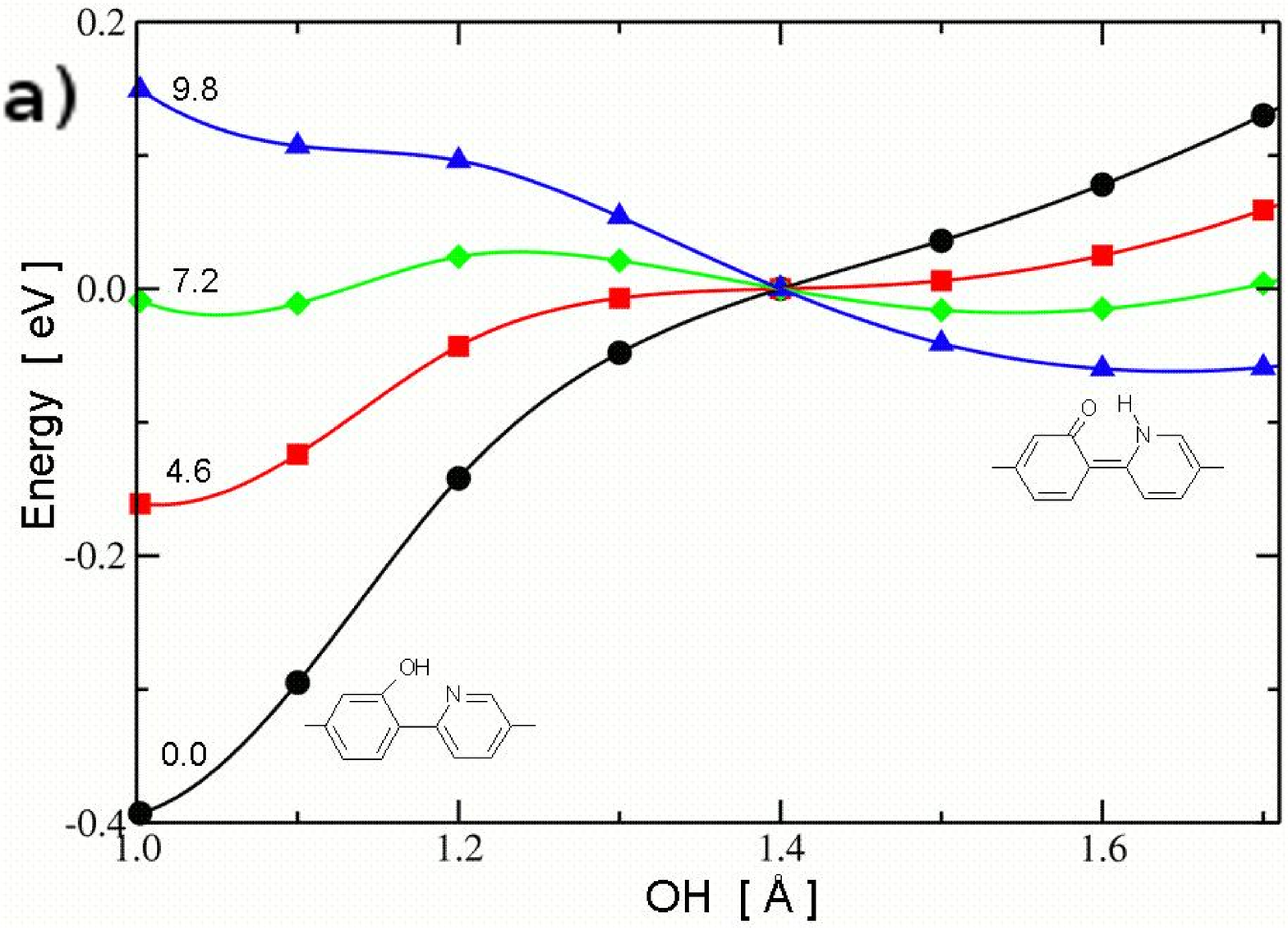}\vspace{3mm}
\includegraphics[scale=0.27,angle=0.0]{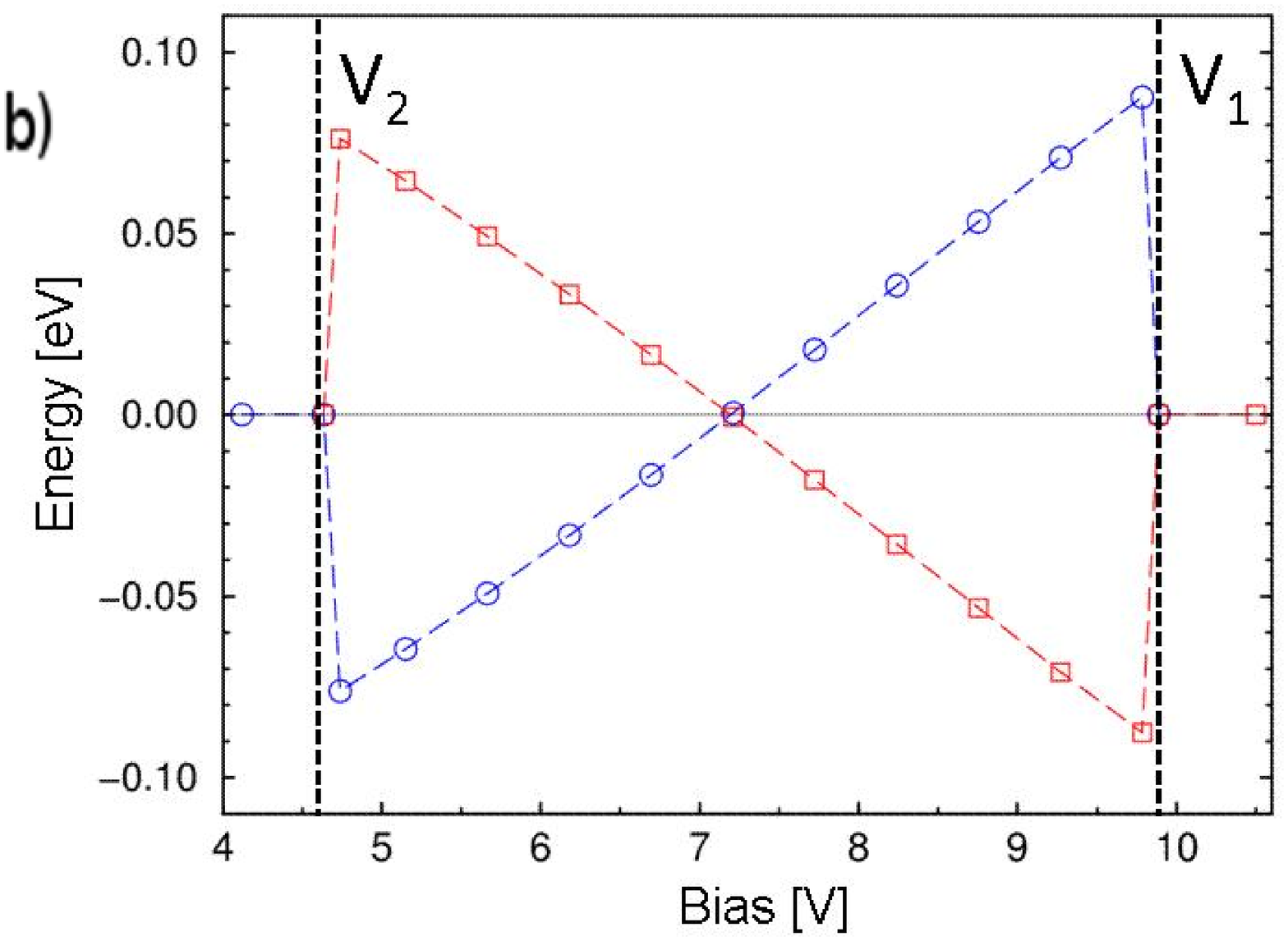}\vspace{3mm}
\includegraphics[scale=0.22,angle=0.0]{F3c.eps}
\caption{Scheme of the switching effect: (a) potential-energy profiles
for the PT reaction at different values of bias in Volts 
indicated above the corresponding MEP, (b) energetic hysteresis
for the switching between the enol- and keto- forms,
(c) model I-V curve of the field-switches – the hysteresis with switching
voltages, V$_1$ and V$_2$, corresponding to the energy barrier-less transitions
between the tautomeric forms: enol-keto and keto-enol, respectively.
The curves (a) and (b) are calculated with B3LYP/cc-pVDZ method for the 2PPy molecule 
terminated with the methyl groups.}
\end{figure}

\begin{figure*}[ht]
\includegraphics[scale=0.9,angle=0.0]{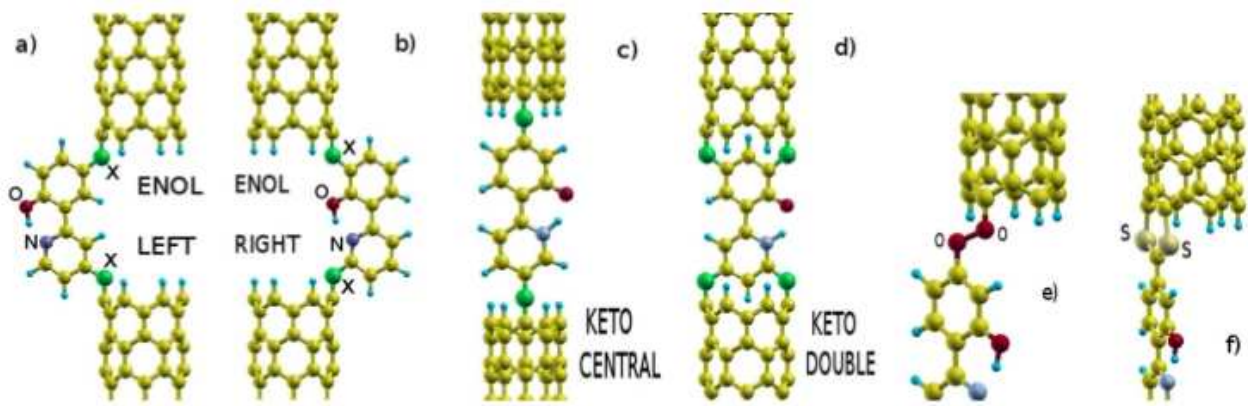}
\caption{Structures of the field-switches with:
enol moieties (a,b), keto moieties (c,d), single-branch current channels (a-c),
double-branch channels (d). X denotes contacts described in the text.
Panels (e) and (f) show the geometries of the -OO- and -CS$_2$- (or -NS$_2$-) contacts,
respectively.}
\end{figure*}



The electric field applied to the molecule may tune the shape 
of the energetical landscape of the ground-electronic state and thus, 
it may change the relative stability of the two tautomeric forms of the molecule. 
One form which is the most stable without the E-field may become less stable 
or even unstable in the higher voltage values which favor another tautomeric form. 
In the studied case the enol form is the only stable form when no filed is applied, 
(see Fig 3a, black filled circles). To stabilize the keto form one must apply 
the electric field along the main exes of the molecule, which is also the direction 
of the proton transfer between the oxygen and nitrogen atoms which 
transforms enol into the keto form. Thus, the distance between 
the oxygen and hydrogen atoms, R(OH), is the best choice of a  driving coordinate 
for an illustrative description of the process. To characterize the process 
in more detail, we construct the minimum-energy profile (MEP) 
along the driving coordinate. 
Along the MEP, all the intramolecular nuclear degrees of freedom are optimized for 
a given frozen value of the driving coordinate R(OH).
Separate ground-state MEPs are calculated for different values of the electric field 
applied to the molecule. Bias values in V are given in Fig. 3a above 
the corresponding MEP additionally distinguished with shape and color 
Those indicate changes in the ground-state energy landscape:
the heights of the enol-to-keto and keto-to-enol tautomerization-reaction barriers 
and the stabilization energy of both forms. MEPs plotted in Fig. 3a represent 
relative energy normalized at R(OH)=1.4 \AA.

Potential-energy profiles sketched in Fig.~3a result in a 'hysteresis' of the field-induced 
switching between the tautomeric forms of the 2PPy molecule terminated with the methyl groups.
The ‘hysteresis’ curve presented in Fig. 3b represents
minimum-energy of a given tautomeric form normalized vs.
the mean energy of both tautomeric forms (the zero-energy of the scale)
at the applied value of bias.
One can see that both tautomeric forms of 2PPy can coexist within a given 
range of the applied voltage,
the terminal values of which define the limits V$_1$ and V$_2$, where the PT process 
enol-to-keto and keto-to-enol takes place, respectively.
The switching values of the voltage impose the limits on the I-V characteristics 
computed for a given tautomeric form (Fig.~3c).
The operation mode of the device is as follows: we start from the enol form,
and when the voltage is increased to the value higher than V$_1$, 
the molecule switches to the keto form. Then, the bias is decreased down to the second 
switching voltage V$_2$, and the reversed transformation to the enol form occurs.
This effect shows up as a jump of the current flowing via the molecule (Fig. 3c).

Similarly to the photo-switches discussed above, also for the field-switches, 
different geometric connections to the CNT can be realized;
they are presented in Fig.~4a-4d. 
In Fig.~4e, we show the zigzag geometry of the peroxy (-O-O-) contact, 
in contrast to the linear geometry of the acetylenic (-C$\equiv$C-) one.
The contact via the CS$_2$ planar group is depicted in Fig.~4f. It is
interesting to explore the CS$_2$ group, since it has been announced as very 
low-resistive contact for gold leads, characterized by low electronic  
work-function.\cite{b20,b43}
To complete these studies, we add also the 
-NS$_2$- connection to our list (in the same geometry as the -CS$_2$- contact).
It has been experimentally evidenced that the field-switching is a robust property of a molecule, 
and not a stochastic phenomenon.\cite{b44}
We show in this work, that the switching effect is not only a property of the
molecule but also strongly depends on the chosen contact.
As one may notice inspecting Fig.~4, the single-branched
contacts might be arranged to the 2PPy switch in the three-fold manner as:
'left', 'right'- or 'central' configuration. The double-branched contacts were
studied earlier by us and showed the negative differential resistance,\cite{b32}
which in some cases can exclude the system as a candidate for a switch.

\begin{table}
\caption{The enol to keto (E-K) tautomeric switching bias V$_1$ [in V]  - 
when the voltage increases - and keto to enol (K-E) switching bias V$_2$
[in V] – when the voltage decreases.}
\begin{tabular}{lcccc}
\hline \\[-0.1cm]
 &  V$_1$  &  V$_2$  &  V$_1$  &  V$_2$ \\
contacts & E-K & K-E & E-K & K-E \\ \hline \\[-0.1cm]
 & \multicolumn{2}{c}{double-branch} & \multicolumn{2}{c}{single-branch central} \\
bare molecule$^{(1)}$  &    &    &  6.2  &   2.6  \\
-CH$_2$- & 6.6 & 2.0 & 9.7 & 4.3 \\
-NH- & 5.4 & 1.4 & 10.3 & 5.4 \\
-O-  & 7.2 & 2.9 & 10.1 & 5.3 \\
-S- & 6.2 & 2.1 & 8.7 & 4.9 \\
direct$^{(2)}$ & & & 8.9 & 3.1 \\
-C$\equiv$C- &  & &  15.9 & 5.6 \\
-O-O- &  &  &  11.5  &  5.7 \\
-CS$_2$-$^{(3)}$ &  &  &  $>$15  & $>$15 \\
-NS$_2$-$^{(3)}$ &  &  &  $>$15  & $>$15 \\
\hline \\[-0.1cm]
 & \multicolumn{2}{c}{single-branch left} & \multicolumn{2}{c}{single-branch right} \\
-CH$_2$- & 8.6 & 3.7 & 7.1 & 2.0 \\
-NH- &  7.9 & 4.2 &  6.3 &  1.4 \\
-O-  &  7.7 & 4.1 &  7.7 & 2.4 \\
-S- & 8.0 & 3.4 & 7.5 & 2.1 \\
\hline \\[-0.1cm]
\multicolumn{5}{l}{$^{(1)}$ molecule is 'free-standing', not connected to leads} \\
\multicolumn{5}{l}{$^{(2)}$ molecule is directly connected to leads } \\
\multicolumn{5}{l}{$^{(3)}$ the enol to keto transition was not found up to 15V} \\
\hline
\end{tabular}
\end{table}

\begin{figure}[ht]
\includegraphics[scale=0.43,angle=0.0]{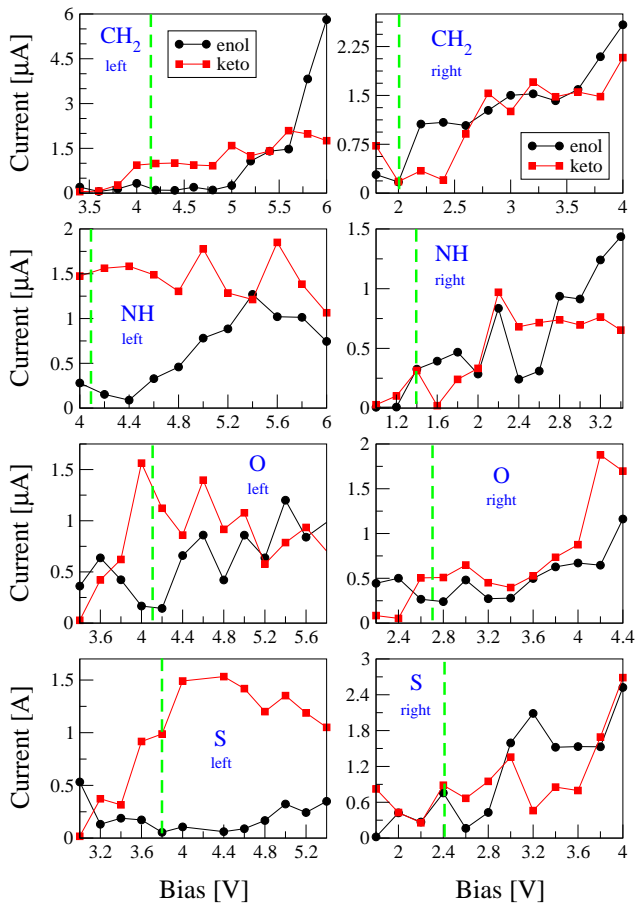}
\caption{The I-V characteristics of the single-branch (left- and right-side)
current-channels. CH$_2$, NH, O and S were chosen for contacts X.
Dashed vertical lines denote the bias at which the lower (V$_2$) tautomeric
transitions occur; this is the measured current jump.} 
\end{figure}

\begin{figure}[ht]
\includegraphics[scale=0.43,angle=0.0]{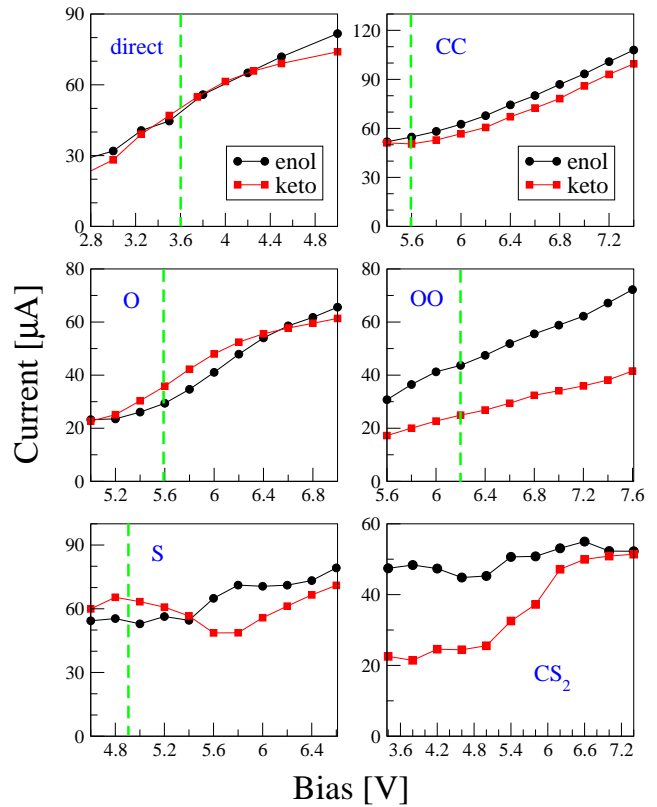}
\caption{The I-V characteristics of the single-branch central current-channels.
For the contacts X, the direct connection, and the atomic contacts:
-C$\equiv$C- (CC), -O-, -O-O- (OO), -S- and -CS$_2$- were chosen. Dashed vertical line
denotes the bias at which the lower transitions (V$_2$) occur;
the measured current jump.}
\end{figure}

In Table~1, we collect the high, V$_1$, and the low, V$_2$, transition voltages
determined for various contacts and structures presented in Fig.~4a-4d.
For completeness, we also show the tautomeric switching
biases for the bare molecule - not connected to the leads.
Inspection of these results leads to the following conclusions:
(i) the operating voltage range for a
given tautomeric form, V$_1$-V$_2$, is relatively wide (several volts),
(ii) various contacts and structures introduce large
diversification of the work regimes.
What is very important, the relatively large switching bias in some cases
is perhaps too high for technological applications. It is generally
more practical, for stability reasons, to utilize the device operating under a
lower transition (V$_2$). Although, in reality the theoretically predicted demands are
overestimated. The calculated bariers would be much smaller when the temperature,
vibrational, interface and tunneling effects were included.\cite{AJ}
Analyzing the results obtained for different connections presented in Table~1,
one may state that the double-branch and 'right'-single-branch structures seem to be the
best promising choice (smaller V$_2$).
While the larger V$_2$ values are given for the 'left'-single-branch connections, and
the single-branch 'central' connections are characterized by the highest switching biases.
On the other hand, as we will see hereafter, the latter case is the only non-NDR case
which is stable with respect to switching.

The I-V curves computed for single-branch side connections, 'left' and 'right', 
are depicted in Fig.~5. These curves correspond to the structures shown in Fig.~4a-4d.
The NDR effect seen in some cases is very often detected in the triangular lattices,
such as graphene and some organic molecules.\cite{b45} Here, we observe the NDR for
the single-branch 'left'- and 'right'-side structures (Fig.~5), and this effect is even 
stronger than for the double-branch case.\cite{b32}
This is generally an undesired effect with respect to the switching performance. 
Some remedy for this
problem possibly could be found using longer anchor groups, 
since a suppression of the quantum conductance oscillations
(leading also to the NDR effect) was reported.\cite{b46}
On the other hand, the energetic barriers to be conquered in the course of
the PT process are not very high (c.f. Fig.~3a) and the vibrational or thermal effects 
may lower them, as mentioned earlier. Thus, moving the transitions to the next bias-point 
(V$_1$ to lower, and V$_2$ to higher) narrowing the hysteretic loop and shifting 
the operation region to the non-NDR part.
The work regimes, chosen for the lower transitions at V$_2$, are marked in 
the I-V characteristics by dashed green lines. 
Despite the NDR effect existing in all cases in Fig.~5, the -NH- and -S- 
'left'-side contacts seem to be promissing for technological applications,  
due to a large gear of the enol and keto currents in the vicinity of 
the switching bias.

Finally, the single-branch central structures were analyzed with respect to their
I-V properties; the results are presented in Fig.~6. One can notice,
upon inspection of the figure, that the keto to enol switching voltage is the
lowest for the case of the direct connection to the leads. Moreover, for the direct
contact, as well as for the acetylenic (-C$\equiv$C-) and oxygenic (-O-) bridges,
the I-V curves of two tautomers are almost identical to each other.
It is not surprising result, since for the high voltage 
the following observations are true: 
\begin{itemize}
\item the range of the quantum-conductance
function entering the current equation (see methods section) is large, 
\item the projected density of states (PDOS) of the O-H-N group in the enol and keto structures
lay closer to the Fermi level than the voltage range which enters the I-V formula.
\end{itemize}
Therefore, all details differentiating one tautomer from the other are summed-up in the
transmission function to the similar values of the current.
If the operating voltage range falls within the range where the characteristic parameters
of the tautomers differ, then the current curves of the tautomers also differ.

\begin{figure*}[ht]
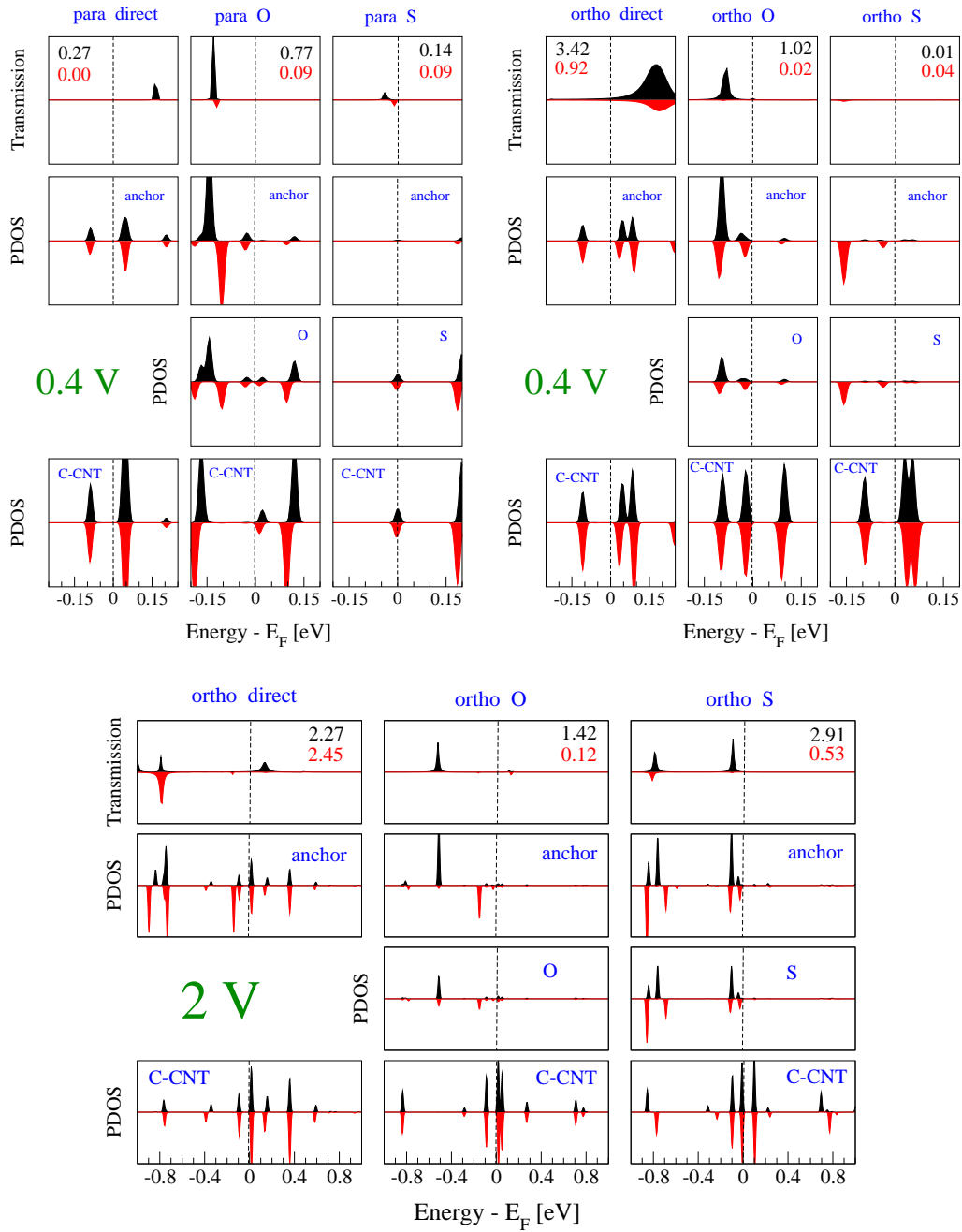

\centerline{
\includegraphics[scale=0.35,angle=0.0]{F7a.eps}\hspace{5mm}
\includegraphics[scale=0.35,angle=0.0]{F7b.eps}}
\vspace{5mm}
\includegraphics[scale=0.4,angle=0.0]{F7c.eps}
\caption{Transmission functions of the photo-switches with the direct,
-O- and -S- contacts, obtained at bias 0.4V (upper panels) and 2V (lower panels); 
the enol is marked in black upside and keto in red downside. 
Numbers are values of the current [in $\mu$A] at 0.4V and 2V, respectively, from  Fig.~2.
PDOS of the $2p$-states of the anchor CC-group, the contacts and the CNT terminal
carbons connected to the contacts.}
\end{figure*}

\begin{figure}[ht]
\includegraphics[scale=0.31,angle=0.0]{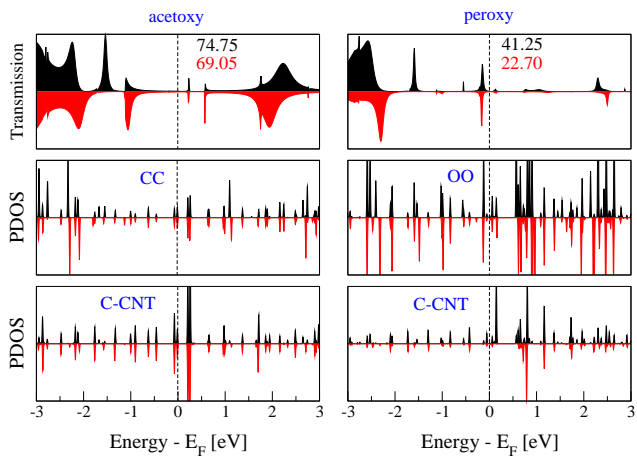}
\caption{Transmission functions of the field-switches in the single-branch
central geometry with the contacts -C$\equiv$C- (CC) and -O-O- (OO), obtained at 
bias 6V; the enol is marked in black upside and keto in red downside. 
Numbers are values of the current [in $\mu$A] at 6V from Fig.~6.
PDOS of the $2p$-states of the CC and OO contacts and the CNT terminal carbons
connected to the contacts.}
\end{figure}

The most important result of this section is the conclusion that for the peroxy (-O-O-) 
contact, a good current-gear for the switching at high voltage ($\sim$5V) is obtained. 
It is not advantageous property of the peroxy 
contacts that they are more resistive than the acetylenic ones (-C$\equiv$C-).
But the peroxy contacts strengthen the diversity of the conductances
of different molecular structures - and this is what we need.

Since the sulfur contact is an electron-rich and highly polarizable moiety, the existence
of undesired NDR effect in this case is not surprising for larger biases.\cite{b45}
We obtained nicely separated enol and keto current
curves for -CS$_2$- at voltages between 3.5V and 5.5V. It is of little practical use, however, 
due to the fact that the switching voltage with increasing bias (V$_1$) is for this contact
higher than 15V - up to the checked PES (see Table~1); and we did not observe switching to
the keto structure. Also the current values for the enol and keto forms show
a tendency to join each other at biases higher than 6V.

\subsection{Transmission and the projected density of states}

The electrical conductivity can be derived from the orbitals hybridization of  
the conducting molecule, atomic contacts and the leads - those orbitals, which
energetically lay close to the Fermi level.
Therefore, plots of the PDOS tell us about reasons for the existence of 
the quantum conductance peaks or their absence. Of course, not all states visible 
on the PDOS plots contribute to the current. In our systems,
there are the lone electron pairs and the molecular orbitals orthogonal to
the conduction direction. On the other hand, it is trivial that the lack 
of states in some energetic region results to no current. 

In Figs. 7 and 8, we plot the quantum conductance (QC) and PDOS of  
photo- and field-switches for selected contacts, for which the I-V curves 
were presented in Figs. 2 and 6.
The electric fields at which the QC and PDOS plots have been obtained correspond
to the bias 0.4V and 2V for the photo-switches (in Fig. 7) 
and 6V for the field-switches (in Fig. 8). 
For convenience, the corresponding values of the current [in $\mu$A] 
are written in the top panels.  

Transition voltage is the bias at which the current starts to grow from zero.\cite{b42}
In the case of photo-switches operating at very low bias regime, the working voltages
of the device are in the range where the enol and keto forms show 
different transition voltages.
The lowest transition biases are for the direct contact,  
medium for oxygen and higher for the sulfur contacts. 
Following the discussion by Wu et al.,\cite{b42} the largest transition voltages
are obtained for those atomic contacts whose $2p$-states lay higher in 
the energy above the Fermi level.
This is also true for the photo-switches and the PDOS presented in Fig.~7. 
The position of the $2p$-states of the anchor group and the contact
and the terminal carbon of the CNT lead is the highest in a case of the sulfur contact.
For the direct contacts, these states are the most close to the Fermi level of all
studied connections. The above sequence corresponds to the obtained current gears.
The states which originate from the lonely pairs, or the orbitals non-overlapping 
with the conduction axis, do not contribute to the quantum conductance.

As for the resistance related problems, it has been stressed in many reports - 
for instance in the recent work by Schulz et al.\cite{b20} -
that the Fermi level mismatch of the conductor and leads is responsible for the low
conductance. Actually, for a good the switch, one tautomeric form should be very resistive 
and the other well conducting in the device-operating bias region.
In contrast to the bias regime of 0.4V (upper panels), for 2V (lower panel)
the best performing contact is that with sulfur, and the next of oxygen. 
The low transition bias of the direct contacts does not work for the switch
in the higher voltage range. 
This is because all PDOS diversities of the enol and keto forms
are summed-up to similar values when the bias range is larger.
Instead, for the oxygen and sulfur contacts, not all PDOS mismatches between two tautomers 
appear in the bias range of 2V.

For the field-swithes operating at higher voltages - around 5-6V - the situation 
is more difficult. One needs to find such contacts, that the nonsymmetric contributions   
to the PDOS from the tautomers do not sum-up in the QC to similar values of current.
In a case of the -C$\equiv$C- contacts, the quantum conductance spectra of tautomers
are nonsymmetric but sum to the similar currents (see Fig.~8).  
On the other hand, the nonsymmetric QC spectra of the peroxy contacts 
give a good current gear. After inspection of the PDOS in the range of [-3,3] eV,
the intensities of the acetoxy states are much smaller than the states of the peroxy case. 
Therefore, the tautomeric diversity is strengthened in the peroxy-contacts case, and they
are the only contacts - of all studied here - which could work for the field-switches. 

Since the energetic positions of the contact states are very important, 
we emphasize that the simple DFT method does not reproduce them well.
One should go beyond this approximation, and for instance the self-interaction corrections
(ASIC) and the constrained DFT (CDFT) method have been applied for the effect 
of the hydrogenetation on the sulfur contacts in the BDT-gold molecular junction.\cite{b47}
On the other hand, our leads and conductors are mainly carbonic, and not metallic.
In contrast to the large underestimation of the energy gap in diamond by  
simple DFT scheme, the carbonic conductive systems, such as graphene and carbon nanotubes,
are well described within this method - the same as the band structures of simple metals.
The DFT failur very strongly shows up when the metallic and non-metallic elements are combined, 
and the localized atomic-like $d$-states (or $f$-states) form.
The relative position of these metallic
states, with respect to the $sp$-states of the non-metallic atoms, is usually placed
too shallow below the Fermi level (or the highest occupied molecular orbital, HOMO, level).
Similarly, the corresponding energies of the unoccupied metallic-states reside too low above the
conduction band minimum (or the lowest unoccupied molecular orbital, LUMO, level).
Such situation leads to large overestimation of the calculated conductance. The systems
studied in this work are purely non-metallic, therefore, they are free of the above described
mismatch of the two kinds of the energy levels. 

Since in this work, we studied an impact of the geometry on the current, another interesting 
phenomenon could be addressed in a future work - namely, the current-driven geometric
and electronic structure rearrangements.
These effects would lead to the non-linear conduction characteristics and 
different switching properties. Such studies within the time-dependent Ehrenfest formalism
were reported for the metallic clusters.\cite{b48}

\section{Conclusions}

We searched for electronic contacts which strengthen the switching properties 
of the molecular devices based on the intramolecular proton transfer process.
Two classes of model molecular photo- and field-switches, operating at lower- and 
higher-bias work regime, respectively, were analyzed.
By means of the first-principles electronic calculations, we found the electronic structure,
Wannier functions, and the electronic transport properties (the I-V curves). 

The larger difference between the currents of the swapped structures,
the better is the performance of a switch. 
Moreover, we would like to have low-resistive contacts, but 
the mechanism of the current on-off switching with applied bias 
and the rules governing the resistance stay in the opposition to each other.
It is demonstrated that for very low- and higher-bias work regimes,
the efficient contacts are chosen according
to different effects: (i) the transition voltage for the current-on
switch and (ii) the resistive conformational selectivity, respectively. 
The low-bias operating range ($\sim$0.5V) of photo-switches 
prefers the direct connection with leads, which is low-resistive.
For the middle voltage work regime (1-2V), 
the oxygen or sulfur are the best atomic contacts.
The high-bias operating range ($>$4V) of the field-switches choses the peroxy contacts.

In addition, we found that the transition voltages of the field-switches
are very sensitive to the choice of contacts. This is caused mainly 
by differences in the polarizabilities of the atomic contacts. \\

\section{Methods}

We performed the density-functional theory\cite{b49} calculations of the molecules 
between the CNT leads using the plane-wave Quantum ESPRESSO code.\cite{b50}
These self-consistent runs were repeated for many discrete
external electric fields with a step visible in the reported figures. 
Then the obtained Bloch functions were used
to generate maximally localized Wannier functions (MLWF)\cite{b51,b52} employing 
the wannier90-2.0.0 code.\cite{b53}
The same package was then utilized for the transport calculations within the 
Landauer-B\"uttiker scheme.\cite{b54}
The obtained quantum conductance (or transmission) for each electric field, 
T($\varepsilon$;E$\sim$V), was embedded in the equation\cite{b55} 
\begin{equation}
I(V) = \int [f(\varepsilon - \varepsilon_F + V/2)  - f((\varepsilon - \varepsilon_F - V/2)]
T(\varepsilon;E) d\varepsilon \nonumber
\end{equation}
with the Fermi-Dirac distribution f($\varepsilon$). This way, we reach the current I 
at the bias V which is equivalent to the
external electric field E assumed in the DFT calculations.

To estimate the voltage values necessary to switch the transistor forms, 
the enol to the keto and back to the enol, we optimized the geometry of the molecule at 
discrete values of the applied electric field (with the step of 0.001 a.u.).
The DFT method implemented in TURBOMOLE [57] was used in
these calculations along with the B3LYP functional and correlation-consistent 
valence double-zeta atomic basis set with polarization functions for all 
atoms (cc-pVDZ)\cite{b57} for thoroughly studied exemplar PPy molecule 
terminated with methyl groups. For all other PPy derivatives the default 
def-SV(P) basis set was used in search of their characteristic V1 and V2 
switching bias values. 

As for the plane-wave calculations setup, the BLYP functional was used. 
The energy cutoff for the plane-waves in the
Quantum Espresso code was set to 30 Ry, and the pseudopotentials were of 
the Martin-Trouliers type from the Fritz-Haber Institute library.
In the quantum conductance calculations, each lead was built by the two CNT
units, and the conductor region included the molecule with the contacts and one 
CNT ring saturated with hydrogens on the left- and right- lead side.
The figures representing the molecular geometry have been
prepared with the XCrySDen package.\cite{b58} \\

{\bf Acknowledgements} \\

This work has been supported by the research projects of the National Science Centre 
of Poland, Grant No. 2012/04/A/ST2/00100.
Calculations were done in the Interdisciplinary Centre of Mathematical and Computer
Modeling (ICM) of the University of Warsaw within the grants G47-7, G56-32 and G29-11.

\end{document}